\documentclass[12pt]{article}

\renewcommand{\baselinestretch}{1.7}

\usepackage{graphicx}
\usepackage{multirow,latexsym,amsfonts,cite}
\usepackage{amsmath}
\usepackage{amssymb}
\usepackage{epsfig, comment}
\usepackage{latexsym}
\usepackage{color}

\newcommand{\be}{\begin{equation}}
\newcommand{\ee}{\end{equation}}
\newcommand{\bea}{\begin{eqnarray}}
\newcommand{\eea}{\end{eqnarray}}

\newcommand{\vect}[1]{\vec{#1}}

\begin{document}
\begin{titlepage}
\null
\begin{flushright}
\end{flushright}

\vskip 1.2cm

\begin{center}
 
 {\Large \bf Magnetic Monopoles in Noncommutative Space-Time: \\
  Second Order of Perturbation}

\vskip 2.0cm
\normalsize

 {\large Miklos L\aa ngvik\footnote{miklos.langvik@helsinki.fi} and Tapio Salminen\footnote{tapio.salminen@helsinki.fi}}

\vskip 0.5cm

  {\large \it Department of Physics, University of Helsinki, \\
              P.O. Box 64, FIN-00014 Helsinki, Finland}

\vskip 1cm

{\bf Abstract}

\end{center}

\renewcommand{\baselinestretch}{1.5}\selectfont

\noindent We investigate the validity of the Dirac Quantization Condition (DQC) for magnetic monopoles  in noncommutative space-time using an extension of the method used by Wu and Yang.
We continue the work started in \cite{1storder} where it was shown that the DQC can be kept unmodified in the first order of the perturbative expansion in the noncommutativity parameter $\theta$. Here we include second order corrections and find that,
in order to find solutions to the noncommutative Maxwell's equations described by the $U_\star(1)$ group, the DQC needs to be modified by perturbative corrections that introduce a dependence on space-time points. Thus the DQC fails to be a topological property of noncommutative space-time.
We comment on the possible origin of this difference.

\end{titlepage}

\newpage

\section{Introduction \label{intro}}

In  \cite{Dirac},  Dirac showed for the first time that the existence of a magnetic monopole would imply the quantization of electric charge.
Today there are a number of subjects, such as non-Abelian monopoles, topology, duality and even quark confinement that are or have been considered in relation to the work of Dirac and the DQC 
\be
{2ge\over \hbar c} = \mathrm{integer} = N.
\label{Diraccond1}
\ee
One of the cornerstones in this development has been the work by Wu and Yang \cite{WuYa}, where it was shown that the DQC can be derived by a method based on singularity-free gauge transformations.
The advantage of this approach is that it directly addresses the question of the singularities that are present in the original derivation of Dirac and that presumably held back the interest in the study 
of magnetic monopoles for some decades after Dirac released his work in 1931.

In this paper we study the validity of the DQC in noncommutative space-time using the Wu and Yang approach\footnote{For a motivational introduction see \cite{1storder}.}. 
We use the noncommutative space defined by the commutator
\be
[\hat{x}^\mu ,\hat{x}^\nu]=i\theta^{\mu \nu}\\, \label{nccom}
\ee
where $\theta^{\mu\nu}$ is a constant antisymmetric matrix. 
Due to the well-known ambiguities arising when time is noncommutative, we only consider the so-called space-space noncommutativity, i.e. $\theta^{0i}=0$, in this work.
The commutation relation \eqref{nccom} is implemented by use of the Weyl-Moyal star product and we take into account that the
gauge group $U_\star(1)$ is non-Abelian due to the star product \cite{Haya}. 
The perturbative expansion of the star product we will be concerned with in this work is given by:
\be
(\Phi \, \Psi) (x)\rightarrow \left(\Phi \star \Psi\right)(x) \equiv
\left[\Phi(x)e^{\frac{i}{2}\theta_{\mu \nu} \label{star}
\frac{\overleftarrow\partial}{\partial x _\mu}
\frac{\overrightarrow\partial}{\partial y_\nu}}
\Psi(y)\right]_{x=y}\,.
\ee
Since we will be working outside of singularities and with continuous functions, we are free to use \eqref{star} which only is defined for smooth functions.

The paper is organized as follows:
In section \ref{repstar} we review the modified version of the Wu and Yang method \cite{WuYa}, originally presented in \cite{1storder}, that takes into account the noncommutativity of space-time.
In section \ref{Gauge} we study the noncommutative gauge transformations to second order in $\theta$.
In section \ref{2nd} and \ref{2order} we define and look for solutions to the noncommutative Maxwell's equations to second order in $\theta$. In section \ref{2order} we arrive at a mathematical inconsistency of the noncommutative gauge transformations and the noncommutative Maxwell's equations under the requirement that the DQC remains topologically valid up to second order in $\theta$. Section \ref{Source} is devoted to a discussion of the two possible noncommutative sources that we may use and section \ref{conclrem} contains our concluding remarks.

\section{Wu-Yang procedure in noncommutative space-time \label{repstar}}

To check the validity of the Dirac quantization condition in noncommutative space-time, we will use a slightly modified version of the original Wu-Yang procedure presented in \cite{WuYa}. 
A more thorough presentation of the modified method is given in \cite{1storder}. 

The procedure is as follows: We will look for a potential in each hemisphere, $A_\mu^N(x)$ and $A_\mu^S(x)$, such that
\begin{enumerate}
\item{The potentials are gauge transformable to each other in the overlapping region of the potentials. For the non-Abelian group $U_\star(1)$ this means that we require
\be \label{gaugetransform}A_{\mu}^{N/S} (x)\rightarrow U(x)\star A_{\mu}^{N/S}(x)\star U^{-1}(x) - iU(x)\star\partial_{\mu}U^{-1}(x)=A_{\mu}^{S/N}(x)\,.\ee}
\item{Both potentials satisfy Maxwell's equations with an appropriate source for the magnetic charge.}
\item{The potentials remain singularity free in their respective regions of validity. That is, Maxwell's equations are solved in such a way that noncommutativity does not produce 
new singularities into the potentials.}
\end{enumerate}

We shall treat the problem as a perturbation series up to second order in $\theta$. In the notation used,  the noncommutative gauge field $A_\mu$ is expanded as $A_\mu = A^0_\mu+A^1_\mu+A^2_\mu+O(\theta^3)$. Here the upper index corresponds to the order in $\theta$ for each correction. In this notation the gauge transformation parameter is (symbolically) expanded as $\lambda =\lambda^0+\lambda^1+ \lambda^2+O(\theta^3)$. To preserve the DQC we require that the $\theta$-corrections to $\lambda$ can be put to zero (or a constant), i.e. $\lambda =\lambda^0+C$, while satisfying the three above requirements. This requirement is the same as in \cite{1storder} and 
we retain it because the interest in the DQC is due to it being independent of space-time points. If the DQC depends on space-time points, the charge of the monopole becomes observer dependent and one could not from the experimental finding of one monopole infer that electric charge is quantized. In our case, higher order
corrections of $\theta$ to the DQC necessarily bring about a dependence on space-time points, as is clear already from dimensional arguments. Our main result is that it is not possible to find solutions when $\lambda^1=0$ and we thus conclude that the DQC will need to be modified thus destroying its topological nature.

We begin by calculating the finite gauge transformations of the fields to second order in $\theta$.

\section{Noncommutative gauge transformations}\label{Gauge}

The noncommutative gauge transformations for the group $U_\star(1)$, with group elements $U^{-1}(x) = e_{\star}^{i\lambda(x)} = 1 + i\lambda(x) + {i^2\over 2!}\lambda(x)\star\lambda(x) + \dots$, are given by
\be
A_{\mu}(x) \rightarrow U(x)\star A_{\mu}(x)\star U^{-1}(x) - iU(x)\star\partial_{\mu}U^{-1}(x). \label{GT}
\ee
There will be $\theta$-contributions from the non-trivial gauge group element $U(x)$ as well as the $\star$-products between the factors of \eqref{GT}. 
The noncommutative gauge transformations can be calculated using the gauge group element calculated to second order in Appendix A with the result \eqref{estar} and the first order 
contribution calculated in \cite{1storder}. Together they give the gauge group element to second order in $\theta$ as
\be
e_{\star}^{i\lambda}  =e^{i\lambda} 
+ \frac{\theta^{ij}\theta^{kl}}{8} e^{i\lambda} \partial_j \partial_l \lambda \left( \frac 12  \partial_i \partial_k \lambda + \frac i3  \partial_i \lambda \partial_k \lambda\right)\ + \mathcal{O}(\theta^3). \label{keke}
\ee
With the help of the gauge group element \eqref{keke} we can calculate the gauge transformation to second order in $\theta$ as is done in Appendix A. The result is, where the first order contributions are calculated in \cite{1storder}
\bea
A_i^0(x) &  \rightarrow & A_i^0(x) + \partial_i\lambda\,, \label{crap23} \\
A_i^1(x) &  \rightarrow & A_i^1(x) +  \theta^{kl}\partial_k\lambda\partial_lA_i^0(x) + {\theta^{kl}\over 2}\partial_k\lambda\partial_l\partial_i\lambda\,, \label{crap24} \\
A_i^2(x) &  \rightarrow & A_i^2(x)+\theta^{kl}\partial_k \lambda \partial_l A^1_i  - \frac 12 \theta^{kl}\theta^{pq}\biggl( \partial_k A^0_i\partial_p\lambda \partial_q\partial_l\lambda - \label{crap25} \\
& & \partial_k \partial_p A^0_i\partial_q\lambda \partial_l\lambda +\frac 13 \left(\partial_k \partial_p\lambda  \partial_l\lambda\partial_q\partial_i\lambda 
- \partial_k \lambda\partial_p\lambda  \partial_l\partial_q\partial_i\lambda \right)\biggr) \,. \notag
\eea
To conclude, due to the first requirement in section \ref{repstar}, we use equations \eqref{crap23}, \eqref{crap24} and \eqref{crap25} to require that the following equations hold:
\bea
 A_i^{N_0}(x) & = & A_i^{S_0}(x) + \partial_i\lambda \,, \label{1stord} \\ 
 A_i^{N_1}(x) & = & A_i^{S_1}(x) +  \theta^{kl}\partial_k\lambda\partial_lA_i^{S_0}(x) + {\theta^{kl}\over 2}\partial_k\lambda\partial_l\partial_i\lambda \,, \label{2ndord} \\
 A_i^{N_2}(x) & = & A_i^{S_2}(x)+\theta^{kl}\partial_k \lambda \partial_l A^{S_1}_i  - \frac 12 \theta^{kl}\theta^{pq}\biggl( \partial_k A^{S_0}_i\partial_p\lambda \partial_q\partial_l\lambda - \label{3rdord}  \\
& &\partial_k \partial_p A^{S_0}_i\partial_q\lambda \partial_l\lambda +\frac 13 \left(\partial_k \partial_p\lambda  \partial_l\lambda\partial_q\partial_i\lambda - \partial_k \lambda\partial_p\lambda  \partial_l\partial_q\partial_i\lambda \right)\biggr) \ . \notag
\eea
Next we will move on to consider the second requirement of section \ref{repstar}, i.e. that the potentials satisfy Maxwell's equations.
 
\section{Noncommutative Maxwell's equations in second order \label{2nd}} 

In noncommutative space Maxwell's equations for a static monopole are given by:
\begin{align}
\epsilon^{\mu\nu\gamma\delta} D_\nu\star\mathcal{F}_{\gamma\delta} &= 0\,,  \label{theother} \\
D_\mu\star\mathcal{F}^{\mu\nu} &= J^\nu \,, \label{sourceeqn}
\end{align}
where $\mathcal{F}_{\mu\nu}= {1\over 2}\epsilon^{\mu\nu\gamma\delta}F_{\gamma\delta}$ is the dual field strength tensor.
As in \cite{1storder} we shall refer to \eqref{theother} as ``Amp\`{e}re's law'' and \eqref{sourceeqn} as ``Gauss's law'' in analogy with electrodynamics.
The noncommutative $U_\star(1)$ field strength tensor and the covariant derivative are given by
\begin{align}
F_{\mu\nu} &= \partial_\mu A_\nu - \partial_\nu A_\mu - ie[A_\mu , A_\nu]_{\star} \,, \\
D_\nu &= \partial_\nu - ie[A_\nu,\cdot]_{\star} \,.
\end{align}
We shall look at the equations \eqref{theother} and \eqref{sourceeqn} as a perturbative series in $\theta$ and check whether we can find solutions for them order by order. For the source we have $J^i=0$, and $J^0 \equiv \rho(r) = 4\pi g\delta(r) + \rho^{1}(r) + \rho^2(r) + \mathcal{O}(\theta^3)$, where the superscript denotes the order in $\theta$. We also note that 
due to the form of equation \eqref{sourceeqn}, we have the additional constraint $D_{\mu}\star J^{\mu} = 0$ on the source. This is however identically satisfied as $A_0 = 0$ and time derivatives vanish in our static case with no 
electric fields present. We may also note that one can define gauge invariant quantities within this formulation that correspond to 
the noncommutative electric and magnetic fields \cite{Gross2, 1storder}. However, as we shall not use them in this work, this note only serves to tell  
that the minimal noncommutative extension of Maxwell's equations considered here contains analogs of the gauge invariant commutative electric and magnetic fields. 

In \cite{1storder} we calculated the noncommutative Maxwell's equations up to 1st order in $\theta$. Here we only quote the result:
\begin{align}
& \left\{ \begin{array}{l}
(\nabla \times B^0)^i = 0\,, \\
(\nabla \times B^1)^i = -\theta^{\gamma\delta} \left[ \partial_j(\partial_\gamma A_0^i \partial_\delta A_0^j) + \partial_\gamma A_j^0 \epsilon^{ijk}\partial_\delta B^0_k   \right]\,,
\end{array} \right. \label{Ampere} \\
& \left\{ \begin{array}{l}
\nabla \cdot \vect{B}_0 = -4\pi\delta(r)\, ,\\
\nabla \cdot \vect{B}_1 = -\rho^1(x) \,.
\end{array} \right.
\end{align}
Here, the notation $F_{\mu\nu}(A^n) = \partial_\mu A_\nu^n - \partial_\nu A_\mu^n$, where $n$ denotes the order of the term in $\theta$ and 
$\vect{B}_n = \nabla\times \vect{A}_n$ is introduced.  Using $\nabla^2 \vec{B} = \nabla(\nabla \cdot \vec{B})+\nabla \times (\nabla \times \vec{B}) $ these combine to give
\be\label{MaxComb}
(\nabla^2 B_1(A_1))^i =-\partial^i \rho^1-\theta^{pq} \{\epsilon^{ijk} \partial^l(\partial^p A_0^k  \partial^j\partial^q A_0^l)- 2\partial^m (\partial^p A^m_0 \partial^q B_0^i ) - \partial^m (\partial^p B_0^m \partial^q A^i_0) \}\,.
\ee
The solutions to this equation were calculated in \cite{1storder} and will be given and used in section \ref{2order}. 

Next, we expand the equations \eqref{theother} and  \eqref{sourceeqn}  to second order in $\theta$. We note that the equations need to be satisfied order by order and so we only need to consider the terms $\mathcal{O}(\theta^2)$. In addition it will help significantly to note that the second order contribution to a star commutator vanishes:
\begin{align}\label{commvanish}
[f,g]_\star &= [f,g] + i\theta^{\mu\nu}\partial_\mu f \partial_\nu g + \frac{1}{2} \left(\frac{i}{2}\right)^2 \theta^{\mu\nu}\theta^{\alpha\beta}(\partial_\mu \partial_\alpha f \partial_\nu\partial_\beta g-\partial_\mu \partial_\alpha g \partial_\nu\partial_\beta f) + \mathcal{O}(\theta^3) \notag\\
&= [f,g] + i\theta^{\mu\nu}\partial_\mu f \partial_\nu g+ \mathcal{O}(\theta^3)\,.
\end{align}

\vspace{5pt}

\noindent \large {\bf Amp\`{e}re's law:} \normalsize We start from Amp\`{e}re's law given by:
\begin{align}
D_\nu \star F^{\mu\nu} &= \partial_\nu F^{\mu\nu}- i[A_\nu, F^{\mu\nu}]_\star \\
&=\partial_\nu F^{\mu\nu}(A^0)+\partial_\nu F^{\mu\nu}(A^1)+\partial_\nu F^{\mu\nu}(A^2)
-i[ A_\nu^0+ A_\nu^1+ A_\nu^2 , F^{\mu \nu}(A^0) + F^{\mu \nu}(A^1) + F^{\mu\nu}(A^2) ]_\star \notag \\
&=\partial_\nu F^{\mu\nu}(A^0)+\partial_\nu F^{\mu\nu}(A^1)+\partial_\nu F^{\mu\nu}(A^2) + i[A^0_\nu, F^{\mu\nu}(A^0)]_\star +i[A^0_\nu, F^{\mu\nu}(A^1)]_\star +i[A^1_\nu, F^{\mu\nu}(A^0)]_\star \,. \notag
\end{align}
All other $\star$-commutators on the second line vanish due to \eqref{commvanish}. 

When we use the zeroth and first order equations \eqref{Ampere} we will only be left with the second order terms
\begin{align}
D_\nu \star F^{\mu\nu} = &\partial_\nu(\partial^\mu A_2^\nu-\partial^\nu A_2^\mu) + \theta^{\gamma\delta} \partial_\nu \left( \partial_\gamma A^\mu_1\partial_\delta A^\nu_0 + \partial_\gamma A^\mu_0\partial_\delta A^\nu_1 \right) \ \\
&+\theta^{\gamma\delta} \{ \partial_\gamma A_0^\nu \partial_\delta(\partial^\mu A_1^\nu-\partial^\nu A_1^\mu) + \partial_\gamma A_1^\nu \partial_\delta(\partial^\mu A_0^\nu-\partial^\nu A_0^\mu) + \theta^{\alpha\beta}\partial_\gamma A^\nu_0\partial_\delta(\partial_\alpha A_0^\mu \partial_\beta A_0^\nu) \} \,. \notag
\end{align}
As before, only the spatial coordinates give a non-zero contribution. From here we can solve for $(\nabla \times B_2)^i \equiv \epsilon^{ijk}\partial_j B_2^k$:
\begin{align}
(\nabla \times B_2)^i = -\theta^{pq} &\{\partial_j(\partial_p A^i_1\partial_q A^j_0 + \partial_p A^i_0\partial_q A^j_1) + \partial_p A^0_j \partial_q(\partial^i A_1^j-\partial^j A_1^i) \notag \\
&+ \partial_p A^1_j \partial_q(\partial^i A_0^j-\partial^j A_0^i) + \theta^{kl}\partial_p A_j^0\partial_q(\partial_k A_0^i \partial_l A_0^j) \}\,.
\end{align}

\vspace{5pt}

\noindent \large {\bf Gauss's law:} \normalsize The calculation of Gauss's law $D_\mu\star\mathcal{F}^{\mu\nu} = J^\nu $ is simplified by $J^i = 0$. Considering only the second order terms we have:
\begin{align}\label{gausss}
-\rho^2 =& \frac{1}{2}\epsilon^{ijk} D^{i}\star F^{jk}(A_0) + \frac{1}{2}\epsilon^{ijk} D^{i}\star F^{jk}(A_1) +\frac{1}{2}\epsilon^{ijk} D^{i}\star F^{jk}(A_2) \notag \\
=& \, \frac{1}{2}\epsilon^{ijk} \theta^{pq}\left[ \partial_p A^i_1 \partial_q (\partial^j A_0^k - \partial^k A_0^j) + \partial_p A^i_0 \partial_q (\partial^j A_1^k - \partial^k A_1^j) + \theta^{rs}\partial_p A^i_0\partial_q(\partial_r A^j_0 \partial_s A^k_0) \right] \notag\\
&+ \frac{1}{2}\epsilon^{ijk} \epsilon^{ljk}D^{i}B^l_2 +\frac{1}{2}\epsilon^{ijk} \partial^i \theta^{pq} (\partial_p A_1^j \partial_q A_0^k +\partial_p A_0^k \partial_q A_1^j ) \,.
\end{align}
The last term can be rewritten as $ \epsilon^{ijk} \partial^i \theta^{pq}\partial_p A_1^j \partial_q A_0^k = \theta^{pq}\left( \partial_p B_1^k \partial_q A_0^k- \partial_p A_1^j \partial_q B_0^j\right)$. This cancels the first two terms in \eqref{gausss} and we get the second order correction to Gauss's law ($\frac{1}{2}\epsilon^{ijk} \epsilon^{ljk}D^{i}B^l_2 = \partial^i B^i_2$):
\begin{align}
\nabla \cdot \vec{B}_2 =& -\rho^2+ \frac{1}{2}\epsilon^{ijk}\theta^{pq}\theta^{rs}\partial_p A^i_0\partial_q(\partial_r A^j_0 \partial_s A^k_0)  \,.
\end{align}
We further note that the second term is zero by a permutation of the indices and we thus have simply:
\begin{align}
\nabla \cdot \vec{B}_2 = -\rho^2\,.
\end{align}
Similar results for the Maxwell's equations have been derived in \cite{Jiang}, but the rest of the calculation and especially the conclusions of \cite{Jiang} differ significantly from ours.

\vspace{5pt}

\noindent \large {\bf Combining the Amp\`{e}re and Gauss laws:} \normalsize We can now combine the results as we did in first order in \eqref{MaxComb}:
\begin{align}
(\nabla^2 B_2)^m =& \partial^m (\nabla \cdot B_2) + (\nabla \times (\nabla \times B_2))^m \notag\\
=& -\partial^m\rho^2 -\epsilon^{mni}\partial^n \theta^{pq} \biggl( \partial_j(\partial_p A^i_1\partial_q A^j_0 + \partial_p A^i_0\partial_q A^j_1)\notag \\
&\hspace{2.2cm}+ \partial_p A^0_j \partial_q \epsilon^{ij l}B^l_1 + \partial_p A^1_j \partial_q \epsilon^{ij l}B^l_0 + \theta^{kl}\partial_p A_j^0\partial_q(\partial_k A_0^i \partial_l A_0^j) \biggr)\,. \label{b2eqn}	
\end{align}

\section{DQC in second order}\label{2order}

In the explicit calculation we will need the original (zeroth order) potentials from \cite{WuYa}, which in cartesian coordinates are given by:
\begin{align}
A_1^{N_0}  = & {-y (r - z)\over (x^2 + y^2) r},\quad A_2^{N_0}  =  {x (r - z)\over (x^2 + y^2) r},\quad A_1^{S_0}  =  {y (r + z)\over (x^2 + y^2) r},\quad \;A_2^{S_0}  =  {-x (r + z)\over (x^2 + y^2) r},\label{potstuff}\\
& \hspace{2.5cm} A_3^{N_0}  =  A_3^{S_0} = A_0^{N_0} = A_0^{S_0} = 0. \nonumber
\end{align}
As before we will take as an assumption that the DQC remains topologically unmodified, i.e. that $\lambda = \lambda^0+\lambda^1+\cdots= \lambda^0 = \frac{2ge}{\hbar c}\phi$, where $\phi = \arctan\left(\frac{x}{y}\right)$.

In the following we shall choose the noncommutative plane as $\theta = \theta^{12}=-\theta^{21}$, other components are set to zero.
With this choice the solutions to the first order Maxwell's equations \eqref{MaxComb} for the potential differences $A^{N_1}-A^{S_1}$  were derived in \cite{1storder} and are given by:
\bea
A^{N_1}_1 - A^{S_1}_1 & = & {2\theta yz(2(x^2+y^2)+z^2)\over (x^2+y^2)^2r^3}\,, \label{solnA1} \\
A^{N_1}_2 - A^{S_1}_2 & = & -{2\theta xz(2(x^2+y^2)+z^2)\over (x^2+y^2)^2r^3}\,, \label{solnA2}\\
A^{N_1}_3 - A^{S_1}_3 & = & 0 \,. \label{solnA3}
\eea
To make a comparison between the Maxwell's equations \eqref{b2eqn} and the gauge transformation \eqref{3rdord} in second order we need to solve for the potentials $A^{N_1}$ and $A^{S_1}$ (not just their difference) in first order, because this quantity appears in \eqref{3rdord}. These also were derived in \cite{1storder} and are given by:
\bea
A^{N_1}_1 & = & \theta\Big({-2x\arctan({x\over y})\over (x^2+y^2)^2}+{y\over 4}\Big[{7\over r^4} - {2\over (x^2+y^2)r^2} + {4z(x^2+y^2+r^2)\over (x^2+y^2)^2r^3}\Big]\Big)\,,  \label{a1whole1} \\
A^{N_1}_2 & = & -\theta\Big({2y\arctan({x\over y})\over (x^2+y^2)^2}+{x\over 4}\Big[{7\over r^4} - {2\over (x^2+y^2)r^2} + {4z(x^2+y^2+r^2)\over (x^2+y^2)^2r^3}\Big]\Big)\,,  \label{a1whole2} \\
A^{N_1}_3 & = & 0. \label{a1whole3}
\eea
From these potentials it is straightforward to obtain the expression for $A^{S_1}_i$, using \eqref{solnA1}, \eqref{solnA2} and \eqref{solnA3}.

In second order we can write down the Maxwell's equations for the separation $B^{N_2}-B^{S_2}$ from \eqref{b2eqn}. They are given by
\bea
\nabla^2(B^{N_2} - B^{S_2})_1 & = & {4{\theta}^2xz\over (x^2+y^2)^3r^{10}}\Big[-375(x^2+y^2)^3 + 131z^2(x^2+y^2)^2 -2z^4(x^2+y^2) -4z^6\Big]\,,  \hspace{20pt} \label{nabnmbsx} \\ 
& & - \partial_1\rho^{N_2} + \partial_1\rho^{S_2}, \nonumber \\
\nabla^2(B^{N_2} - B^{S_2})_2 & = & {4{\theta}^2yz\over (x^2+y^2)^3r^{10}}\Big[-375(x^2+y^2)^3 + 131z^2(x^2+y^2)^2 -2z^4(x^2+y^2) -4z^6\Big]\,,  \label{nabnmbsy} \\
& & - \partial_2\rho^{N_2} + \partial_2\rho^{S_2}, \nonumber \\
\nabla^2(B^{N_2} - B^{S_2})_3 & = & {4{\theta}^2\over (x^2+y^2)^4r^{10}}\Big[120(x^2+y^2)^5-900(x^2+y^2)^4z^2-1285(x^2+y^2)^3z^4  \label{nabnmbsz}\\
& & -1289(x^2+y^2)^2z^6-652(x^2+y^2)z^8-132z^{10}\Big] - \partial_3\rho^{N_2} + \partial_3\rho^{S_2}. \nonumber
\eea

These equations are difficult to solve analytically. Fortunately this is not needed, since we only want to compare these equations to the ones coming from the gauge transformation in the overlap of the potentials \eqref{3rdord}. To find out what happens to the DQC in this second order calculation, we must compare them to the equations resulting from the gauge transformations in order to satisfy criterion 1 and 2 of section \ref{Gauge}. They are given by
\begin{align}
\nabla^2(B^{N_2} - B^{S_2})_1^{GT} & = {4 {\theta}^2xz\over (x^2 + y^2)^3 r^{10}} \Big(-321(x^2 + y^2)^3 + 205(x^2 + y^2)^2z^2 + 26(x^2 + y^2)z^4 + 4z^6\Big)\,, \hspace{25pt} \label{nabnmbsxgt} \\
\nabla^2(B^{N_2} - B^{S_2})_2^{GT} & = {4 {\theta}^2yz\over (x^2 + y^2)^3 r^{10}} \Big(-321(x^2 + y^2)^3 + 205(x^2 + y^2)^2z^2 + 26(x^2 + y^2)z^4 + 4z^6\Big)\,, \label{nabnmbsygt} \\
\nabla^2(B^{N_2} - B^{S_2})_3^{GT} & = {4 {\theta}^2\over (x^2 + y^2)^4r^{10}}\Big(144(x^2 + y^2)^5 -564(x^2 + y^2)^4z^2 - 455(x^2 + y^2)^3z^4 \label{nabnmbszgt} \\
&  - 403(x^2 + y^2)^2z^6 -188(x^2 + y^2)z^8 - 36z^{10}\Big)\,. \nonumber
\end{align}
These are the equations resulting from the gauge transformation of the second order contribution of the potential \eqref{3rdord} using the potentials \eqref{potstuff} and \eqref{a1whole1}-\eqref{a1whole3}, after
first taking the curl and then the Laplace-operator of the resulting transformation. 

In order for the DQC to be fulfilled, the equations \eqref{nabnmbsx}-\eqref{nabnmbsz} and \eqref{nabnmbsxgt}-\eqref{nabnmbszgt} need to be satisfied simultaneously. We may simplify this system of equations by subtracting \eqref{nabnmbsx} from \eqref{nabnmbsxgt}, \eqref{nabnmbsy} from \eqref{nabnmbsygt} and \eqref{nabnmbsz} from \eqref{nabnmbszgt}. The resulting system of equations is given by
\bea
-\partial_x(\rho^{N_2}-\rho^{S_2}) & = & {8 {\theta}^2xz\over (x^2 + y^2)^3 r^{8}} \Big(27(x^2 + y^2)^2 + 10(x^2 + y^2)z^2 + 4z^4\Big)\,,  \label{GTMEOMX} \\
-\partial_y(\rho^{N_2}-\rho^{S_2}) & = &   {8 {\theta}^2yz\over (x^2 + y^2)^3 r^{8}} \Big(27(x^2 + y^2)^2 + 10(x^2 + y^2)z^2 + 4z^4\Big)\,,  \label{GTMEOMY} \\
-\partial_z(\rho^{N_2}-\rho^{S_2}) & = &   {2 {\theta}^2\over (x^2 + y^2)^3 r^{8}} \Big(48(x^2 + y^2)^4 + 624(x^2 + y^2)^3z^2 + 1036(x^2 + y^2)^2z^4 \label{GTMEOMZ} \\
& & + 736(x^2 + y^2)z^6 + 192z^8\Big) \,. \nonumber
\eea
We can then differentiate equation \eqref{GTMEOMX} with respect to $y$ and equation \eqref{GTMEOMY} with respect to $x$ and perform a subtraction between the two:
\be
0=(\partial_x\partial_y-\partial_y\partial_x)(\rho^{N_2}-\rho^{S_2})  =  0, \label{soufunc}
\ee 
where the $0$ on the left hand side is due to the partial derivatives commuting and the $0$ on the right hand side is due to a calculation of the expression $(\partial_x\partial_y-\partial_y\partial_x)(\rho^{N_2}-\rho^{S_2})$ by using 
equations \eqref{GTMEOMX} and \eqref{GTMEOMY}.
 We get the following two additional equations in a similar manner:
\begin{align}
0=({\partial_x\partial_z-\partial_z\partial_x})(\rho^{N_2}-\rho^{S_2})  =  {24 {\theta}^2 x \over (x^2 + y^2)^5 r^8}&\Big(41 (x^2 + y^2)^4 + 426 (x^2 + y^2)^3 z^2 + 704 (x^2 + y^2)^2 z^4 \notag\\
 &+ 496 (x^2 + y^2) z^6 + 128 z^8\Big)\,, \label{soufunc2}  \\
0=({\partial_y\partial_z-\partial_z\partial_y})(\rho^{N_2}-\rho^{S_2})   =  {24 {\theta}^2 y \over (x^2 + y^2)^5 r^8}&\Big(41 (x^2 + y^2)^4 + 426 (x^2 + y^2)^3 z^2 + 704 (x^2 + y^2)^2 z^4 \notag\\ 
&+ 496 (x^2 + y^2) z^6 + 128 z^8\Big)\,. \label{soufunc3} 
\end{align}
In the above we have set the left hand sides of equations \eqref{soufunc}, \eqref{soufunc2} and \eqref{soufunc3} to zero. This can be motivated by the following:
Since the partial derivates of $\left(\rho^{N_2}-\rho^{S_2}\right)$ exist (they are given by \eqref{GTMEOMX}, \eqref{GTMEOMY} and \eqref{GTMEOMZ}), and are continuous outside the origin, the function $\left(\rho^{N_2}-\rho^{S_2}\right)$ itself must be continuous in this region. Thus the partial derivatives commute and we can safely put the left hand sides to zero.

The equations \eqref{soufunc2} and \eqref{soufunc3} have no solution (except when $x = y = 0$) 
and consequently the DQC cannot be satisfied topologically in our 
model. The situation is unlikely to be different had we divided the potential in another way (something other than hemispheres). This seems to be the case as we did not 
introduce any new singularities into the overlap of the gauge potentials with the division into hemispheres and the division of \cite{WuYa} only serves to remove the unphysical singularites (Dirac string singularity in 
the commutative case). 

Thus we can conclude that there does not exist potentials $A^N_\mu$ and $A^S_\mu$ that would simultaneously satisfy Maxwell's equations and be gauge transformable to each other by \eqref{GT}. Therefore we get our final result: The DQC cannot hold topologically in this model.
Obviously this does not mean that we could not introduce a perturbative source for the noncommutative monopole. It only states that one cannot keep the DQC independent of space-time points in this model. 
More interestingly, due to the problems we face in second order, one must in fact make the DQC dependent of space-time points already in first order of $\theta$ to be able to have a 
consistent set of potentials that are gauge transformable to each other in the second order of $\theta$. This follows because the curl of $\partial_i\lambda^2$ in equation \eqref{3rdord} vanishes, and hence it does not contribute 
to the Maxwell's equations in second order. As a consequence, one must include a nontrivial gauge transformation parameter $\lambda^1$ to have a consistent solution of the noncommutative Maxwell's equations. 
Calculating the correction explicitly from equation \eqref{3rdord} should in principle be possible, but has proven technically very challenging. We have achieved our less ambitious goal, however, in proving that the DQC necessarily receives corrections in $\theta$ and therefore no longer remains a topological property of space-time.

It should be mentioned that we could in principle add gauge covariant terms to the source that do not contain a Dirac delta function. These terms could for instance have the form $\eta_i\theta_{kl}F_{kl}\star D_j\star F_{ij}$, where $\eta_i$ is a constant vector and $D_j$ the $U_{\star}(1)$ covariant derivative. This term is possible, since it does not modify the first order equations in the overlap and it is gauge covariant. They can still modify the expression for the potentials \eqref{a1whole1}-\eqref{a1whole3}, and consequently the conclusion. However, we have checked explicitly that a similar contradiction arises even when including such terms and thus they are omitted in favor of a much simpler description. This can further be motivated by the fact that all the terms that we can form using $\eta_i, D_i, A_i(x)$ and $F_{ij}$, do not contain the delta function and are therefore not part of a perturbative deformation of the monopole equations. Since our objective is to study the deformed Maxwell monopole equations, such terms would be \emph{ad hoc} and are thus omitted.


\subsection{Comparison to previous results}

Although a direct generalization of the Wu-Yang formulation of the Dirac monopole, such as here, has not been considered 
previously, there have been many studies of noncommutative BPS-monopoles. The works include 
perturbative studies of the $U_{\star}(2)$  \cite{U2} and $U_{\star}(1)$  \cite{U1} BPS-monopoles, as well as 
nonperturbative
studies of the $U_{\star}(1)$ \cite{nonpert} BPS-monopoles, generalized to other groups in \cite{lecht}. These 
constructions share the assumption that the definition of magnetic charge in the 
BPS-limit
may be taken over, without change, to the noncommutative case. Our result shows that this assumption 
fails in the perturbative case \footnote{Essentially, the assumption taken is that the noncommutative space reduces to a commutative 
space in the $\theta \rightarrow 0$ limit, and that this limit is a physical one (this assumption is needed to be able to speak of magnetic charge). In our view, it is not.
Although the limit is smooth and mathematically well defined in the classical theory, it does not represent a physical process.
That is, if we take our space to be noncommutative, then there is no known physical process that reduces it to a commutative space. Another argument that strongly supports this point of view can be made if we take quantum processes into account.
In this case, the limit $\theta \rightarrow 0$ is actually non-smooth due to the mixing of UV and IR divergences. Since the world is ultimately quantum, this fact should not be neglected. 
This means that $U(1)$ is only an approximate 
symmetry and $U_{\star}(1)$ is the exact symmetry of the noncommutative Maxwells equations. Therefore, if one is to speak of magnetic charge, it must be defined for a $U_{\star}(1)$-theory, not a $U(1)$-theory. 
These remarks have been made specific in \cite{1storder}.}. Therefore it is to be expected that the definition of noncommutative 
magnetic charge in the nonperturbative treatment is also subject to problems. However, this is still an open question 
that we would like to return to in more detail in another work.

As our approach produces a perturbative expansion of the gauge potentials $A^{N}$ and $A^{S}$, it is natural to ask whether it is compatible with the well-known perturbative Seiberg-Witten map \cite{SeiWitt}.
\begin{align}
\hat{A}_i&= A_i-\frac{1}{4}\theta^{kl}\{A_k,\partial_l A_i+F_{li} \}+\mathcal{O}(\theta^2) \label{WittA}\,,\\
\hat{\lambda}&= \lambda +\frac{1}{4}\theta^{ij}\{\partial_i \lambda, A_j \}+\mathcal{O}(\theta^2) \,,
\end{align}
where $\{\cdot, \cdot\}$ is an anticommutator. We have checked explicitly that the first order potentials derived from \eqref{WittA} do not satisfy the equations of motion \eqref{MaxComb}.
This could be due to the physical singularity at the origin, but irrespective of the reason for this failure we note
that, as observed in \cite{SeiWitt}, the map \eqref{WittA} necessarily produces a $\hat\lambda$ that depends on the potential $A_0$. If it did not, the gauge groups $U(1)$ and $U_\star(1)$
in the ordinary and noncommutative theories respectively, would be identical. As this is clearly not the case, any approach based on the perturbative Seiberg-Witten map cannot produce potentials compatible with the DQC.

However, in the case of the nonperturbative Seiberg-Witten map constructed for the U(1)-part of the field-strength \cite{dudes}, the situation may improve considerably. In this case, the Seiberg-Witten map
turns into an expression containing the open Wilson line which has been observed to be important in the case of the noncommutative Aharonov-Bohm effect \cite{ChaLaSaTu}. If we consider the noncommutative space to be a 
deformation of the commutative one and because we know that the $U(1)$-monopole and the Aharonov-Bohm effect are related in the commutative case, one would expect them to be related also in the deformed, noncommutative case. Something they seem not to be, given the perturbative analysis here and the result of \cite{ChaLaSaTu}. Furthermore, due to the analysis of
\cite{oogu}, one may indeed expect that the nonperturbative Seiberg-Witten map would give rise to a different kind of delta function support for the field-strength, thus changing the analysis made here 
and perhaps even the result of the calculation.

     
\section{A noncommutative particle source \label{Source}}

The above calculation shows that, at least in perturbation theory, we cannot force the DQC to hold topologically by choosing a source term with the zero order contribution given by the Dirac delta function.
 However, we might still ask what a possible 
noncommutative particle source might look like, be it a monopole, an electrically charged particle or something else. Naturally, even if we cannot find a DQC supportive source for 
the monopole, it should still be possible to find a source term for e.g. an electrically charged particle, as charged particles must be described somehow in $U_\star(1)$ electrodynamics, 
if the theory is to have any connection with commutative Maxwellian electrodynamics.

To determine a possible noncommutative particle source, we have to discuss the symmetry of the equations.   
Firstly, equation \eqref{sourceeqn} transforms as $U(x)\star D_\mu\star\mathcal{F}^{\mu0}\star U^{-1}(x)$ under gauge transformations on the left-hand side. Namely, it is gauge covariant. Therefore, the source must also transform this way.
Secondly, the left-hand side is $O(1,1)\times SO(2)$ symmetric and consequently, the source must also be that. Thirdly, as a correspondence principle when $\theta \rightarrow 0$, we want to recover the 
Dirac delta-function for the source. We shall take this view here because it will be compatible with the theory we presented, although as is clear, it is rather restrictive from 
a general point of view.

To begin with, any realistic generalization of a particle-source must in the noncommutative Maxwell's equations transform under gauge transformations. 
Therefore extensions of the delta function, such as
\be
\delta_{NC}^3(r) = {1\over \sqrt{(4\pi  \theta)^3}}\exp\big({-r^2\over 4 \theta}\big), \label{plpp}
\ee
must be discarded. They do not contain the gauge potential and therefore do not transform under gauge transformations.

For the consistency of Maxwell's equations we need to find a source that is covariant up to second order of perturbation. We have indeed found two such expansions, which perhaps surprisingly, have all of their coefficients uniquely fixed. The form of the possible sources is thus strongly constrained by gauge covariance.

If we suppose we have a source that transforms gauge covariantly 
in the second order of $\theta$ and we call it $\rho_2$, where $\rho_{NC} = \rho_0 + \rho_1 + \rho_2 + ...$, one can calculate the gauge transformation it must satisfy using the gauge group element \eqref{estar}. It is given by:
\be\label{gaugecovcondition}
\rho_2 \rightarrow \rho_2 + \theta^{ij}\partial_i\lambda\partial_j\rho_1 + {\theta^{ij}\theta^{kl}\over 2}\Big(\partial_k\lambda\partial_i\lambda\partial_j\partial_l\rho_0 - \partial_j\lambda\partial_l\rho_0\partial_i\partial_k\lambda\Big),
\ee
where $\lambda$ is the local gauge group parameter. Equation \eqref{gaugecovcondition} is the gauge covariance requirement in second order of $\theta$ for the second order correction to the source.

If we start with a first order source of the form $\theta^{kl}\partial_k \left(A_{l}\delta^3(r)\right) $\footnote{The first order contribution, up to a sign change, was found in \cite{dqc} and was also considered in \cite{1storder}.} we find 
a gauge covariant source up to second order in $\theta$, satisfying all our symmetry requirements, as:
\begin{align}
\rho =&\rho^{0}+\rho^{1}+\rho^{2}+\mathcal{O}(\theta^3)=\;4\pi g \biggr(\delta^3(r) -\theta^{kl}\partial_k \left(A_{l}\delta^3(r)\right)    \notag\\
+&\theta^{ij} A_j^{1}\partial_i \delta^3(r) +\theta^{ij}\theta^{kl}\left[ A_j^{0}\partial_k\left(\partial_i A_l^0 \delta^3(r) +A_l^{0}\partial_i \delta^3(r)\right) +\frac{1}{2}A_i^{0}A_k^{0}\partial_j\partial_l\delta^3(r) \right] +\mathcal{O}(\theta^3)\biggr)\,. \label{crap}
\end{align}
Due to the requirement of gauge covariance, the first order contribution to the source is unique up to the position of the partial derivative and the numerical coefficient in front. The second order contribution was found by using the most general Ansatz dimensionally possible, performing the transformation according to \eqref{crap23} and \eqref{crap24} and finally comparing with the gauge covariance condition \eqref{gaugecovcondition}. An interesting point is that the second order coefficients as well as the coefficient for the first order term are all uniquely determined by specifying the form of the first order contribution $\theta^{kl}\partial_k \left(A_{l}\delta^3(r)\right) $.

The other first order source term leading to a gauge covariant expansion in second order $\theta$ is $\theta^{kl} A_{l}\partial_k\delta^3(r)$. The expansion is then given by:
\begin{align}
\label{source2}
\rho =\;4\pi g \biggr(\delta^3(r) -\theta^{ij}A^0_{j}\partial_i\delta^3(r)  
-\theta^{ij}A_j^{1}\partial_i\delta^3(r)   + \frac{1}{2}\theta^{ij}\theta^{kl}A_i^{0}A_k^{0}\partial_j\partial_l \delta^3(r)  + \mathcal{O}(\theta^3)\biggr)\,.
\end{align}
The two second order sources \eqref{crap} and \eqref{source2} are the only gauge covariant expansions consistent with the noncommutative Maxwell's equations. In our case,
the inclusion of the source does not change the contradiction in equations \eqref{soufunc2} and \eqref{soufunc3}. This follows because the only contribution that the source \eqref{crap} or \eqref{source2} has on the 
function $B^2_i$ is at the origin $r=0$, where the noncommutativity of space-time makes the theory more singular due to the physical singularity at the position of the monopole. As this point is not included in the zeroth order potentials it follows that it is not included in the full expressions $A^N_i$ and $A^S_i$. Thus we do not need to consider the second order source contribution in the calculation of $B^2_i$.

\section{Discussion and Conclusions \label{conclrem}}

We have used the approach of Wu and Yang in Weyl-Moyal space to show that the Dirac Quantization Condition fails to hold topologically in second order of perturbation in the noncommutativity parameter $\theta$. This failure could be due to the perturbative approach used, as the infinite non-locality induced by the $\star$-product is only apparent in the nonperturbative approach \footnote{A similar perturbative modification of quantization has been noticed for the magnetic flux of vortex solutions in 2+1 dimensional Chern-Simons theory coupled to a scalar field in the BPS setting \cite{CS}.}. It is clear that the topological DQC cannot be recovered with the inclusion of any finite number of $\theta$-corrections, but a final verdict for the DQC cannot be given before a nonperturbative 
treatment of this problem has been accomplished. A full nonperturbative calculation with the method considered here would thus be an interesting continuation of our work.

The most intuitive explanation for our result would be the breaking of rotational invariance. Since rotational invariance is directly related to the fibre bundle
construction of the Wu-Yang potentials in commutative space-time, it may indeed be that the breaking of it leads to a nontopological DQC in noncommutative space-time.

As another possible explanation we should mention the following: NCQED is $CP$-violating \cite{CPT} and it is known that in flat commutative space-time \cite{witt} and 
even in curved space-time \cite{arch}, a $CP$-violating theory  
necessarily leads to the monopole aquiring an electric charge and the failure of the DQC. One could be led to believe that this phenomenon, also called charge dequantization, could 
be exactly what we have observed in this work. At present, the relation of our result to charge dequantization is not well understood. 
We would like to point out however, that the noncommutative Maxwell's equations are manifestly $CP$-violating whereas the $CP$-violation observed in the charge dequantization 
phenomenon of commutative electrodynamics only occurs if one adds extra terms to the free Lagrangian of electrodynamics. This does indicate some difference in the two approaches. 

Assuming that a nonperturbative treatment leads to the same conclusion, it is interesting to speculate over the implications this result might have. First of all, since the charge of the matter fields is quantized in noncommutative theories \cite{Haya, ChaPreSheTu}, 
we are not in need of another explanation for the quantization of charge. Furthermore, since
the Aharonov-Bohm effect can be formulated in a gauge invariant manner in noncommutative quantum mechanics  \cite{ChaLaSaTu}, the noncommutative theories seem to make a 
difference of outcome between the experimentally observed Aharonov-Bohm effect and the DQC.
Thus these two results, closely related in commutative space-time, seem to be unrelated in the noncommutative theory.
As a consequence, one might argue that the lack of observations of magnetic monopoles is related to the deformed structure of
space-time at very small scales. 

\noindent {\large \bf Acknowledgments}

We are especially indebted to Anca Tureanu for useful comments and advice in the preparation of this work. We also acknowledge Masud Chaichian, 
Peter Horv\'{a}thy and Archil Kobakhidze for helpful comments and remarks. 

\section*{Appendix A}

\section*{Gauge transformation in second order of $\theta$}
In this Appendix we calculate the second order contribution to the finite gauge group element and its gauge transformation.
We do this by first calculating the second order terms in $\theta$ for the combination $\lambda(x)\star\lambda(x)$. Then for $\lambda(x)\star\lambda(x)\star\lambda(x)$ and 
then $\lambda(x)\star\lambda(x)\star\lambda(x)\star\lambda(x)$, et.c. The second order terms for $\lambda(x)\star\lambda(x)$ are
\be
{1\over 2!}({i\over 2})^2\theta^{ij}\theta^{kl}\partial_i\partial_k\lambda(x)\partial_j\partial_l\lambda(x).
\ee
For $\lambda_{\star}(x)^3 = \lambda(x)\star\lambda(x)\star\lambda(x)$ they are (remember here that all first order terms are 0)
\bea
\lambda_{\star}(x)^3 & = & \big(\lambda(x)^2 + {1\over 2!}({i\over 2})^2\theta^{ij}\theta^{kl}\partial_i\partial_k\lambda(x)\partial_j\partial_l\lambda(x) \big)\star\lambda(x) + \mathcal{O}(\theta^3) \label{3term} \\
& = & \lambda(x)^3 + {1\over 2!}({i\over 2})^2\theta^{ij}\theta^{kl}\partial_i\partial_k(\lambda(x)^2)\partial_j\partial_l\lambda(x) + {1\over 2!}({i\over 2})^2\theta^{ij}\theta^{kl}\lambda(x)\partial_i\partial_k\lambda(x)\partial_j\partial_l\lambda(x) + \mathcal{O}(\theta^3) \nonumber \\
& = & \lambda(x)^3 + {2\over 2!}({i\over 2})^2\theta^{ij}\theta^{kl}\partial_i\lambda(x)\partial_k\lambda(x)\partial_j\partial_l\lambda(x) + {3\over 2!}({i\over 2})^2\theta^{ij}\theta^{kl}\lambda(x)\partial_i\partial_k\lambda(x)\partial_j\partial_l\lambda(x) + \mathcal{O}(\theta^3). \nonumber
 \eea
For $\lambda_{\star}(x)^4 = \lambda(x)\star\lambda(x)\star\lambda(x)\star\lambda(x)$ we get (from now on we omit the $x$ in $\lambda(x)$)
\bea
\lambda_{\star}^4 & = & \big( \lambda^3 + {2\over 2!}({i\over 2})^2\theta^{ij}\theta^{kl}\partial_i\lambda\partial_k\lambda\partial_j\partial_l\lambda + {3\over 2!}({i\over 2})^2\theta^{ij}\theta^{kl}\lambda\partial_i\partial_k\lambda\partial_j\partial_l\lambda\big)\star\lambda + \mathcal{O}(\theta^3) \label{4term} \\
& = & \lambda^4 + {1\over 2!}({i\over 2})^2\theta^{ij}\theta^{kl}\partial_i\partial_k\lambda^3\partial_j\partial_l\lambda +  {2\over 2!}({i\over 2})^2\theta^{ij}\theta^{kl}\lambda\partial_i\lambda\partial_k\lambda\partial_j\partial_l\lambda + {3\over 2!}({i\over 2})^2\theta^{ij}\theta^{kl}\lambda^2\partial_i\partial_k\lambda\partial_j\partial_l\lambda + \mathcal{O}(\theta^3) \nonumber \\
& = & \lambda^4 +  {8\over 2!}({i\over 2})^2\theta^{ij}\theta^{kl}\lambda\partial_i\lambda\partial_k\lambda\partial_j\partial_l\lambda + {6\over 2!}({i\over 2})^2\theta^{ij}\theta^{kl}\lambda^2\partial_i\partial_k\lambda\partial_j\partial_l\lambda + \mathcal{O}(\theta^3), \nonumber
\eea
where in the first line we used the result \eqref{3term}.
Doing similar calculations for $\lambda_{\star}^5$, $\lambda_{\star}^6$, $\lambda_{\star}^7$ et.c., and collecting terms we get a series that looks like:
\bea\label{seriesexp}
e_{\star}^{i\lambda} & \stackrel{{\theta}^2}{=} & {i^2\theta^{ij}\theta^{kl}\over 8}\Big[\partial_i\partial_k\lambda\partial_j\partial_l\lambda\Big({i^2\over 2!} + {i^3\over 3!}3\lambda + {i^4\over 4!}6\lambda^2 + {i^5\over 5!}10\lambda^3 + {i^6\over 6!}15\lambda^4 + {i^7\over 7!}21\lambda^5 + {i^8\over 8!}28\lambda^6 + \mathcal{O}(\lambda^7)\Big) \nonumber \\
& + & \partial_i\lambda\partial_k\lambda\partial_j\partial_l\lambda\Big({i^3\over 3!}2 + {i^4\over 4!}8\lambda + {i^5\over 5!}20\lambda^2 + {i^6\over 6!}40\lambda^3 + {i^7\over 7!}70\lambda^4 + {i^8\over 8!}112\lambda^5 + \mathcal{O}(\lambda^6)\Big)\Big] . \label{mara}
\eea
The notation $\stackrel{{\theta}^2}{=}$ is exactly like an equal sign, but it only displays explicitly quantities of second order in $\theta$ to the right of the sign. The expression \eqref{mara} may be recognized as the following series
\be
e_{\star}^{i\lambda}  \stackrel{{\theta}^2}{=}  {i^2\theta^{ij}\theta^{kl}\over 8}\sum_{n=1}^{\infty}\Big[\partial_i\partial_k\lambda\partial_j\partial_l\lambda{i^{n+1}\over (n+1)!}\lambda^{n-1}\sum_{s=1}^{n}s + 
2\partial_i\lambda\partial_k\lambda\partial_j\partial_l\lambda{i^{n+2}\over (n+2)!}\lambda^{n-1}\big(\sum^{n}_{t=0}\sum_{s=1}^{n-t}s\big)\Big]\,. \label{crap2}
\ee
Recalling the results 
\bea
\sum_{s=1}^{n}s & = & {n(n+1)\over 2}, \\
\sum_{t=1}^{n}t^2 & = & {n(n+1)(2n+1)\over 6},
\eea
the result \eqref{crap2} may be cast into the form
\be
e_{\star}^{i\lambda}   \stackrel{{\theta}^2}{=} {i^2\theta^{ij}\theta^{kl}\over 8}\sum_{n=0}^{\infty}\Big[\partial_i\partial_k\lambda\partial_j\partial_l\lambda{i^{n+2}\over (n+2)!}\lambda^{n}({n\over 2}+1)(n+1) + 
\partial_i\lambda\partial_k\lambda\partial_j\partial_l\lambda{i^{n+2}\over (n+2)!}\lambda^{n-1}{n(n+1)(n+2)\over 3}\Big]. \label{serie}
\ee
Furthermore, if we look more closely at the terms 
\bea
I & = & \sum_{n=0}^{\infty}{i^{n+2}\over (n+2)!}\lambda^{n}({n\over 2}+1)(n+1) \label{hmm1} \,,\\
I' & = & \sum_{n=0}^{\infty}{i^{n+3}\over (n+3)!}\lambda^{n}{n(n+1)(n+2)\over 3} \label{hmm2}\,, 
\eea
we see that \eqref{hmm1} can be reduced to 
\bea
I & = & {\partial\over \partial\lambda} \sum_{n=0}^{\infty}{i^{n+2}\over (n+2)!}\lambda^{n+1}({n\over 2}+1) = {i\over 2} {\partial\over \partial\lambda}  \sum_{n=0}^{\infty}{i^{n+1}\over (n+1)!}\lambda^{n+1} \\
& = &  {i\over 2}{\partial\over \partial\lambda}\big[  \sum_{n=0}^{\infty}{i^{n}\over n!}\lambda^{n} - 1\big] = -{1\over 2}e^{i\lambda}. \label{kiva1}
\eea
Also, \eqref{hmm2} can be reduced to 
\bea
I' & = & {1\over 3} \sum_{n=0}^{\infty}{i^{n+2}\over (n+2)!}\lambda^{n-1}n(n+1)(n+2) =  {1\over 3}{\partial^3\over \partial\lambda^3}\sum_{n=0}^{\infty}{i^{n+2}\over (n+2)!}\lambda^{n+2} \notag\\
 &=& {1\over 3}{\partial^3\over \partial\lambda^3}(e^{i\lambda} - (1 + i\lambda )) = -\frac{i}{3}e^{i\lambda}\,. \label{kiva2}
\eea
Thus the full expression \eqref{seriesexp} can be written in the simple form:
\be
e_{\star}^{i\lambda}   \stackrel{{\theta}^2}{=} \frac{\theta^{ij}\theta^{kl}}{8} e^{i\lambda} \partial_j \partial_l \lambda \left( \frac 12  \partial_i \partial_k \lambda + \frac i3  \partial_i\lambda \partial_k \lambda\right) \,. \label{estar}
\ee
Since this expression contains only the exponential function and derivatives of $\lambda$ the single-valuedness condition is the same as for $e^{i\lambda}$. This is to be expected since the $\star$ would not touch a constant angle $2 n \pi$ if it is added to the exponential.

We can now write down the expression for the full gauge transformation \eqref{GT}. Taking only the second order terms:
\begin{align}
A^2_i \rightarrow & e_\star^{-i\lambda}\star(A_i^0(x) + A_i^1(x)+ A_i^2(x))\star e_\star^{i\lambda} - ie_\star^{-i\lambda}\star\partial_i e_\star^{i\lambda} + \mathcal{O}(\theta^3) \notag \\
=& A^2_i + \theta^{kl}\partial_k \lambda \partial_l A^1_i+ \left(\frac i2 \right)^2\theta^{kl}\theta^{pq}\left( \frac 12 \partial_p\partial_k \lambda \partial_q\partial_l \lambda A^0_i + 2\partial_p \lambda \partial_q\partial_l \lambda \partial_k A^0_i-2\partial_q \lambda \partial_l \lambda \partial_k\partial_p A^0_i\right) \notag \\
&+ \frac{1}{2}\left( i \partial_k \lambda \partial_p \lambda-\partial_k  \partial_p \lambda\right)\left(i \partial_l\partial_q\partial_i\lambda +i^2\partial_l\partial_q \lambda\partial_i\lambda -i \partial_q \lambda\partial_l\lambda\partial_i\lambda -\partial_l\partial_i\lambda\partial_q\lambda-\partial_l\lambda\partial_q\partial_i\lambda\right) \biggr)\notag \\
&+ \frac{\theta^{kl}\theta^{pq}}{8}\left(A_i^0  \partial_k \partial_p\lambda \partial_l \partial_q\lambda+ \partial_i\lambda\partial_k \partial_p\lambda \partial_l \partial_q\lambda -i \partial_i \left[\partial_q \partial_l\lambda \left(\frac 12 \partial_p \partial_k\lambda + \frac i2 \partial_p \lambda \partial_k\lambda \right)  \right] \right) \notag \\
=& A^2_i + \theta^{kl}\partial_k \lambda \partial_l A^1_i - \frac 12 \theta^{kl}\theta^{pq}\left( \partial_k A^0_i\partial_p\lambda \partial_q\partial_l\lambda - \partial_k \partial_p A^0_i\partial_q\lambda \partial_l\lambda +\frac 13 \left(\partial_k \partial_p\lambda  \partial_l\lambda\partial_q\partial_i\lambda - \partial_k \lambda\partial_p\lambda  \partial_l\partial_q\partial_i\lambda \right)\right)\,.
\end{align}

\renewcommand{\baselinestretch}{1}\selectfont


\begin{thebibliography}{99}


\bibitem{1storder}
  M.~L\aa ngvik, T.~Salminen and A.~Tureanu, 
  Phys. Rev. {\bf D 83} (2011) 085006,  
  [arXiv:1101.4540 [hep-th]].

\bibitem{Dirac}
  P.~A.~M.~Dirac,
  Proc.\ Roy.\ Soc.\ Lond.\ {\bf A 133} (1931) 60.
  
\bibitem{WuYa}
  T.~T.~Wu and C.~N.~Yang,
  Phys.\ Rev.\ {\bf D 12} (1975) 3845.


\bibitem{Haya}
  M.~Hayakawa,
  Phys.\ Lett.\ {\bf B 478} (2000) 394,
  [arXiv:hep-th/9912094].


\bibitem{Gross2}
  D.~J.~Gross, A.~Hashimoto and N.~Itzhaki,
  Adv.\ Theor.\ Math.\ Phys.\  {\bf 4} (2000) 893,
  [arXiv:hep-th/0008075].
    
\bibitem{Jiang}
  L.~Jiang,
  arXiv:hep-th/0001073.



\bibitem{U2}
  D.~Bak,
  Phys.\ Lett.\  {\bf B 471} (1999) 149, 
  [arXiv:hep-th/9910135]; \\
  K.~Hashimoto, H.~Hata, S.~Moriyama,
  JHEP {\bf 9912} (1999) 021, 
  [arXiv:hep-th/9910196]; \\
   S.~Goto, H.~Hata,
  Phys.\ Rev.\  {\bf D 62} (2000) 085022,
  [arXiv:hep-th/0005101]. 

\bibitem{U1}
  K.~Hashimoto, T.~Hirayama,
  Nucl.\ Phys.\  {\bf B 587} (2000) 207, 
 [arXiv:hep-th/0002090]; \\
   L.~Cieri and F.~A.~Schaposnik,
  Res.\ Lett.\ Phys.\  {\bf 2008} (2008) 890916,
  [arXiv:0706.0449 [hep-th]].

\bibitem{nonpert}
   D.~J.~Gross and N.~A.~Nekrasov,
  JHEP {\bf 0007} (2000) 034, 
  [arXiv:hep-th/0005204].\\
   M.~Hamanaka, S.~Terashima,
  JHEP {\bf 0103} (2001) 034,
  [arXiv:hep-th/0010221].  
 
\bibitem{lecht} 
  O.~Lechtenfeld, A.~D.~Popov,
  JHEP {\bf 0401} (2004) 069,
  [arXiv:hep-th/0306263].

\bibitem{SeiWitt}
 N.~Seiberg and E.~Witten,
  JHEP {\bf 9909} (1999) 032,
  [arXiv:hep-th/9908142].

\bibitem{dudes} 
  Y. Okawa and H. Ooguri, 
  [arXiv:hep-th/0104036]; \\
  S. Mukhi and N. V. Suryanarayana, 
  JHEP 0105 (2001) 023, [arXiv:hep-th/0104045]; \\
  H. Liu and J. Michelson, 
  [arXiv:hep-th/0104139].

\bibitem{ChaLaSaTu}
  M.~Chaichian, M.~L\aa ngvik, S.~Sasaki and A.~Tureanu,
  Phys.\ Lett.\ {\bf B 666} (2008) 199,
  [arXiv:0804.3565 [hep-th]].

\bibitem{oogu}
  K.~Hashimoto and H.~Ooguri,
  Phys.\ Rev.\ D\ {\bf 64} (2001) 106005
  [arXiv:hep-th/0105311].

\bibitem{dqc}
  M.~Chaichian, S.~Ghosh, M.~L\aa ngvik and A.~Tureanu,
  Phys.\ Rev.\ {\bf D 79} (2009) 125029,
  [arXiv:0902.2453 [hep-th]].

\bibitem{CPT}
  M.~M.~Sheikh-Jabbari,
  Phys.\ Rev.\ Lett.\  {\bf 84} (2000) 5265,
 [arXiv:hep-th/0001167].

\bibitem{witt}
  E. Witten, Phys. Lett. B {\bf 86} (1979) 283.

\bibitem{arch}
  A.~Kobakhidze and B.~H.~J.~McKellar,
  Class.\ Quant.\ Grav.\  {\bf 25} (2008) 195002,
  [arXiv:0803.3680 [hep-th]].

\bibitem{CS}
  G.~S.~Lozano, E.~F.~Moreno, F.~A.~Schaposnik,
  JHEP {\bf 0102 } (2001)  036,
  [arXivhep-th/0012266];\\
  D.~Bak, S.~K.~Kim, K.~-S.~Soh, J.~H.~Yee,
  Phys.\ Rev.\  {\bf D 64} (2001)  025018,
  [arXivhep-th/0102137];\\
  P.~A.~Horvathy, L.~Martina and P.~C.~Stichel,
  Nucl.\ Phys.\ {\bf B 673} (2003) 301,
  [arXiv:hep-th/0306228].

\bibitem{ChaPreSheTu}
  M.~Chaichian, P.~Pre\v{s}najder, M.~M.~Sheikh-Jabbari and A.~Tureanu,
  Phys.\ Lett.\ {\bf B 526} (2002) 132,
  [arXiv:hep-th/0107037].





\end{thebibliography}
\end{document}